\documentclass[aps,prmaterials,superscriptaddress,reprint]{revtex4-1}
\usepackage[english]{babel}
\usepackage{amsfonts}
\usepackage{amsmath}
\usepackage{graphicx}
\usepackage{mathtools}
\usepackage[caption=false]{subfig}
\usepackage{multirow}

\usepackage{soul}
\usepackage{color}

\newcommand{\C}[1]{$^{\circ}$C}

\begin{document}

\title{Structure and morphology of low mechanical loss TiO$_2$-doped Ta$_2$O$_5$}

\author{Mariana A. Fazio}
\email{Mariana.Fazio@colostate.edu}
\affiliation{Department of Electrical and Computer Engineering and NSF ERC for Extreme Ultraviolet Science and Technology, Colorado State University, Fort Collins, CO, USA}

\author{Gabriele Vajente}
\affiliation{LIGO Laboratory, California Institute of Technology, Pasadena, CA, USA}

\author{Alena Ananyeva}
\affiliation{LIGO Laboratory, California Institute of Technology, Pasadena, CA, USA}

\author{Ashot Markosyan}
\affiliation{Department of Applied Physics, Ginzton Laboratory, Stanford University, Stanford, CA, USA}

\author{Riccardo Bassiri}
\affiliation{Department of Applied Physics, Ginzton Laboratory, Stanford University, Stanford, CA, USA}

\author{Martin Fejer}
\affiliation{Department of Applied Physics, Ginzton Laboratory, Stanford University, Stanford, CA, USA}

\author{Carmen S. Menoni}
\email{Carmen.Menoni@colostate.edu}
\affiliation{Department of Electrical and Computer Engineering and NSF ERC for Extreme Ultraviolet Science and Technology, Colorado State University, Fort Collins, CO, USA}
\date{\today}

% \linenumbers

\begin{abstract}
	
Amorphous oxide thin films play a fundamental role in state-of-the art interferometry experiments, such as gravitational wave detectors where these films compose the high reflectance mirrors of end and input masses. The sensitivity of these detectors is affected by thermal noise in the mirrors with its main source being the mechanical loss of the high index layers. These thermally driven fluctuations are a fundamental limit to optical interferometry experiments and there is a pressing need to understand the underlying processes that lead to mechanical dissipation in materials at room temperature. Two strategies are known to lower the mechanical loss: employing a mixture of Ta$_2$O$_5$ with $\approx$ 20\% of TiO$_2$ and post-deposition annealing, but the reasons behind this are not completely understood. In this work, we present a systematic study of the structural and optical properties of ion beam sputtered TiO$_2$-doped Ta$_2$O$_5$ films as a function of the annealing temperature. We show for the first time that low mechanical loss is associated with a material morphology that consists of nanometer sized Ar-rich bubbles embedded into an atomically homogeneous mixed titanium-tantalum oxide. When the Ti cation ratio is high, however, phase separation occurs in the film which leads to increased mechanical loss. These results indicate that for designing low mechanical loss mixed oxide coatings for interferometry applications it would be beneficial to identify materials with the ability to form ternary compounds while the dopant ratio needs to be kept low to avoid phase separation.

\end{abstract}

\maketitle

\section{Introduction}

Amorphous oxide coatings play a fundamental role in a wide range of optical systems, from high power lasers to quantum circuits. In particular, mixed oxide materials have attracted much attention due to the potential to tune their optical properties by varying the dopant concentration. These coatings have found their way into diverse applications such as laser-damage-resistant coatings \cite{mangote2012femtosecond}, thin film transistors \cite{fortunato2012oxide} and high reflectance mirrors in gravitational-wave detectors \cite{harry2006titania}. In the latter, these mirrors constitute the end masses of the interferometers of both Advanced LIGO \cite{abbott2016gw150914} and Advanced Virgo \cite{acernese2014advanced}. The oxide coatings take on a major role in these state-of-the-art experiments as the mechanical damping in these materials leads to the Brownian motion noise which decreases the sensitivity of the detectors. The thermally driven fluctuations (thermal noise) lead to optical path length variations and are a fundamental limit to many optical interferometry experiments \cite{saulson1990thermal, numata2004thermal} including atomic clocks \cite{ludlow2008sr}. There is a pressing need to understand the underlying processes that lead to mechanical dissipation in these materials. In the case of gravitational wave detectors, the main source of thermal noise is found in the mechanical loss of the high index material in the stack and is currently the ultimate barrier to improve design sensitivity \cite{penn2003mechanical}.

Currently there are two known processes that lead to a reduction in the mechanical losses of high index oxide coatings: post deposition thermal treatment (annealing) and doping. In the first case, annealing has been reported to significantly reduce the mechanical loss of the high index material in the reflective stack in addition to decreasing stress and optical absorption \cite{martin2010effect, netterfield2005low}. In particular for tantala (Ta$_2$O$_5$) films it has been found that annealing has the strongest effect on the reduction of the coating loss angle, even when films are grown at elevated substrate temperatures \cite{vajente2018effect}. In the second case, it was found that by doping tantala with 20-25\% of titania (TiO$_2$) the mechanical losses could be reduced by around 40\% \cite{abbott2016gw150914,granata2016mechanical,granata2019amorphous}.  This led to designing the high reflectance mirrors in both Advanced LIGO and Advanced Virgo to be composed of alternating layers of silica (SiO2) and titania-doped tantala (TiO$_2$:Ta$_2$O$_5$). The thermal noise reduction achieved due to the use of this particular mixed oxide coating has contributed to an increased sensitivity of the observatories which was instrumental to recent major detections \cite{aasi2015advanced}.

The causes of the reduction in mechanical loss achieved with titania doping and the role of the dopant concentration in the material mechanical loss are not yet understood. Harry et. al. evaluated films with different titania concentrations up to 55\% with all coatings subjected to annealing in air at 600\C{}, and reported that the mechanical loss reached a minimum for this specific dopant concentration of around 20\% \cite{harry2006titania}. Several authors confirmed this decrease in mechanical loss as well \cite{granata2016mechanical,granata2019amorphous,flaminio2010study,principe2015material,gras2017audio,amato2018high}.  There are however, only a few studies on the structure and properties of titania-doped tantala films and their relation to mechanical loss. Investigations of structural modifications of titania-doped tantala using transmission electron microscopy by Bassiri et. al showed that Ti doping, in particular cation ratios around 0.2 - 0.3, promotes structural homogeneity at the nearest-neighbor level which seems to correlate with low mechanical loss. In their study the film with the lowest mechanical loss (Ti/Ta cation ratio of 0.283) was also the least oxygen deficient, suggesting that Ti doping might prevent oxygen loss which also contributes to lowering the mechanical loss \cite{bassiri2013correlations}. On follow up work Bassiri et. al. found that annealing of the films did not greatly affect the structure, only subtle changes in the short-order were observed, but proposed that the medium range order could show significant changes as a function of the annealing temperature \cite{bassiri2016order}. Modeling and experiments in zirconia doped tantala have confirmed that modifications in the medium range order that result in more corner sharing tetrahedra favors low room temperature mechanical loss in that mixture \cite{prasai2019high}.

Herein we report the results of a detailed study of the evolution in the atomic structure, morphology, optical properties and mechanical loss of reactively sputtered titania-doped tantala upon annealing. It is shown that for a dopant cation concentration of 0.27, the annealing process induces atomic mixing leading to the formation of a titania-tantala compound, while for a higher cation concentration of 0.53 the film can be regarded as a molecular mixture of titania and tantala. The crystallized phase for the film with the lowest cation concentration is indexed as TiTa$_{18}$O$_{47}$. Instead, the film with the highest dopant concentration is phase separated with only the titania crystallizing at 500\C{}. At an annealing temperature of 600\C{}, where the structural modifications are clearly identified, the titania-doped tantala films with cation ratio of 0.27 achieve the lowest mechanical loss and absorption loss at the laser interferometer wavelength of 1064 nm. At these conditions, the films are less dense due to the presence of Ar-rich bubbles. In combination, these results show, for the first time, that the lowest mechanical loss is associated with a material morphology that consists of nanometer sized Ar-rich bubbles embedded into an atomically homogenous mixed titanium-tantalum oxide, which features reduced optical loss.

\section{Results} \label{sec:results}

\begin{figure}[h!]
	\includegraphics[width=\linewidth]{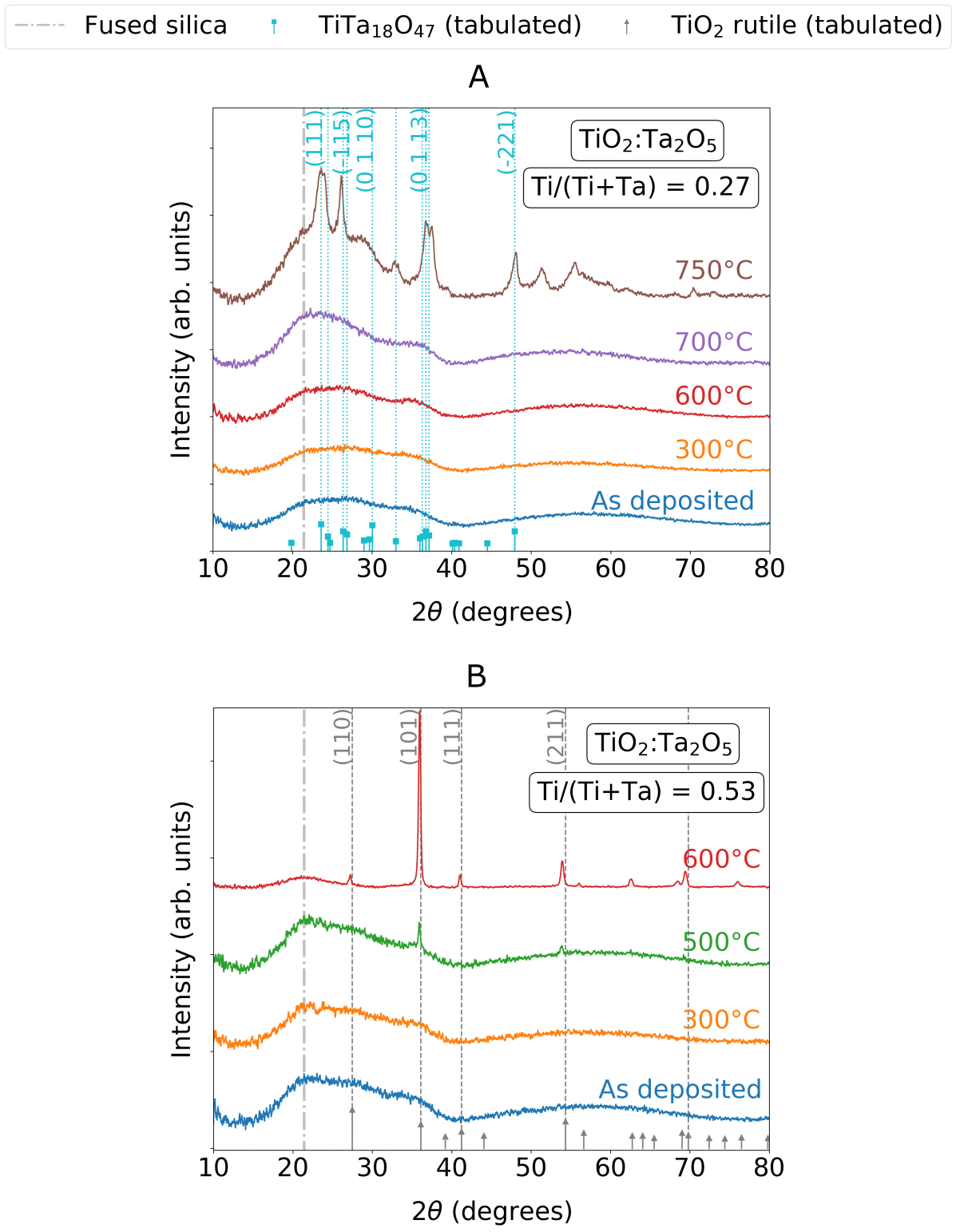}
	\caption{\label{fig:xrd} Diffractograms for (A) titania-doped tantala film with a cation ratio = 0.27 and (B) titania-doped tantala film with a cation ratio = 0.53 as a function of annealing temperature. The fused silica substrate peak position is indicated in (A) and (B) with its main feature being a broad peak centered around 2$\theta$ = 21$^{\circ}$.Tabulated peak positions for the TiTa$_{18}$O$_{47}$ compound and for titania in the rutile phase are included.}
\end{figure}

Grazing incidence x-ray diffraction (GIXRD) results for titania-doped tantala films with different cation ratio concentrations are presented in figure \ref{fig:xrd}.

The as-deposited film with a cation dopant concentration of 0.27 shows an amorphous structure with broad features comparable to those observed for tantala (see Figure 1 of Supplementary Material \cite{supplementary}). These broad features are unchanged up to an annealing temperature of 600\C{}. However, the main broad feature around 2$\theta$ = 16 - 40$^{\circ}$ is sharper and higher in intensity after annealing at 700\C{}, which is indicative of the onset of the film crystallization. In fact, annealing to 750\C{} reveals peaks superimposed over the main broad features which indicate partial crystallization of the film. The peaks of the crystallized structure correspond to TiTa$_{18}$O$_{47}$ (PDF 00-021-1423). This compound was observed in powder samples consisting of a mixture of tantala and titania in varying concentrations \cite{waring1968effect}. These XRD measurements are reported in a limited angular range, 2$\theta$ = 50$^{\circ}$ [23], which do not permit indexing of three peaks observed at higher positions in the diffractograms of figure \ref{fig:xrd}A. The crystal system of TiTa$_{18}$O$_{47}$ was identified as monoclinic. The relative intensities of the peaks match fairly well with the tabulated values except, for a peak expected at 2$\theta$ = 30.03$^{\circ}$ which is not observed. This could be due to the peak being superimposed with the amorphous background still present or could also be indicative of texture in the structure. The cation ratio necessary to form TiTa$_{18}$O$_{47}$ is around 0.05 while the actual ratio of the film is much higher (0.27) so it is reasonable to expect that only a small portion of the amorphous layer has crystallized. Figure \ref{fig:xrd}B presents the GIXRD spectra for the titania-doped tantala film with a dopant cation ratio of 0.53. The films as-deposited and annealed at 300\C{} indicate an amorphous structure with the broad features also observed for pure tantala and the mixed film with a lower dopant cation ratio. For an annealing temperature of 500\C{} two peaks centered at 2$\theta$ = (35.93 $\pm$ 0.03)$^{\circ}$ and 2$\theta$ = (53.9 $\pm$ 0.3)$^{\circ}$ appear, indicating that the film is partially crystallized. For the following annealing temperature of 600\C{} the diffractogram consists mainly of sharp peaks with only a broad feature remaining associated with the substrate. The peaks correspond to the titania rutile structure with a preferential orientation in the [101] direction. There are also significant shifts in peak positions that can be linked to a variation in lattice parameters, as there are no clear indications of the presence of residual stress in the film. In this case, one of the constituents of the mixture, titania, fully crystallizes while the tantala remains amorphous featuring a clear phase separation in the material. Remarkably, annealing this film at temperatures higher than the crystallization temperature of pure tantala (up to 850\C{}) does not induce crystallization of the tantala phase.
 
\onecolumngrid

\begin{figure}[h!]
	\includegraphics[width=0.65\linewidth]{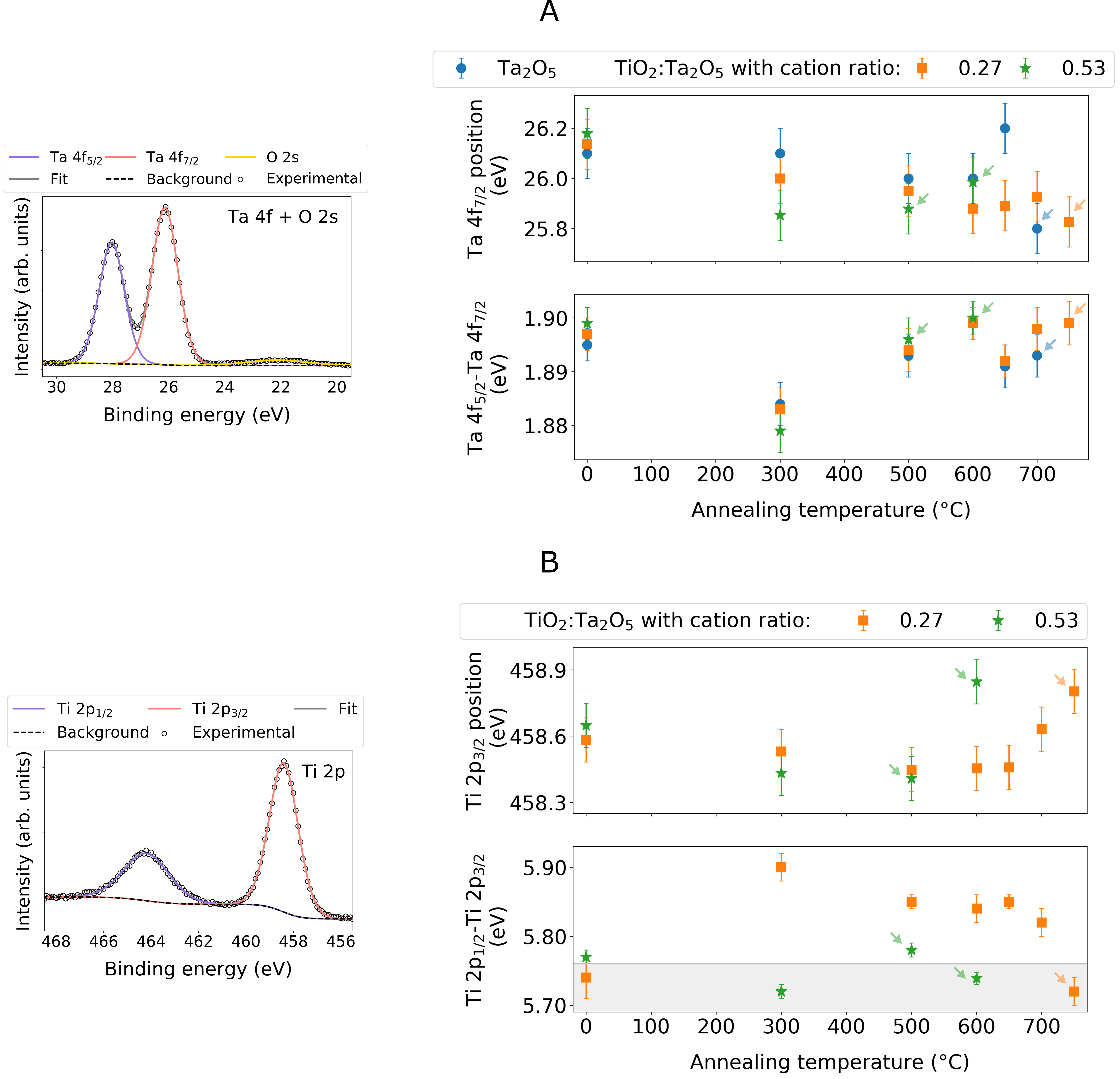}
	\caption{\label{fig:xps} XPS results for titania-doped tantala films. (A) Position of Ta 4f$_{7/2}$ peak and energy separation of Ta 4f doublet. The position and peak separation for a tantala film are shown for reference. (B) position of Ti 2p$_{3/2}$ peak and energy separation of Ti 2p doublet in titania-doped tantala films. The shaded region indicates the tabulated values of energy separation for Ti(IV). Arrows indicate crystallized films as determined by the presence of sharp peaks in the diffractograms.}
\end{figure}

\twocolumngrid

Annealing of the titania-doped tantala films also modified the material structure observed by x-ray photoelectron spectroscopy (XPS) from the evolution of the Ta 4f and Ti 2p peaks. Figure \ref{fig:xps}A shows the variation of the Ta 4f$_{7/2}$ peak position and the binding energy separation between the Ta doublet for tantala and titania-doped tantala films as a function of the annealing temperature. The position of the Ta 4f$_{7/2}$ peak and the energy separation are well within the tabulated ranges for tantala \cite{nist_database,demiryont1985effects,simpson2017xps,atanassova1998x} in all films at the different annealing temperatures. This indicates that the Ta oxidation state is that of tantala and that neither the annealing nor the dopant inclusion induce significant changes in the chemical environment of the Ta atoms.

A similar analysis for the Ti bonding environment in titania-doped tantala films is shown in figure \ref{fig:xps}B. The Ti 2p$_{3/2}$ peak position versus annealing for all films agrees with tabulated values for Ti(IV) \cite{nist_database}. However, the energy separation of the doublet for the film with a cation ratio of 0.27 presents an interesting behavior. For the as deposited film the energy separation is in good agreement with the expected energy separation for Ti(IV). After annealing there is a significant increase of the doublet energy separation. Similar high values of energy separation have been reported for poorly oxidized titania films \cite{haukka1993dispersion,platau1977oxidation}, but in this case the position of the O 1s and Ti 2p$_{3/2}$ peaks do not correspond to a low oxidation state. This indicates that the chemical environment of the Ti atoms changes significantly with annealing up to the crystallization temperature and no longer corresponds to a titania environment, which could be indicative of an atomic mixture. After partial crystallization of the TiTa$_{18}$O$_{47}$ phase the energy separation is again within the range corresponding to Ti(IV), likely due to the fact that a large portion of the material remains amorphous in a tantala and a titania phase. This indicates that crystallization effectively breaks the atomic mixture induced in the material by the annealing process. For the titania-doped tantala film with a cation ratio of 0.53, the energy separation of the doublet components are consistent with Ti(IV) and do not show any marked dependency on the annealing temperature even after complete crystallization of the titania phase at 600\C{}. The energy separation measured after annealing at 500\C{} is only deviated by 1$\sigma$ from the tabulated values, which is not statistically significant. This indicates that the Ti atoms are in a titania chemical environment in the as-deposited films and after annealing, regardless of its structure being amorphous or partially crystallized.

\onecolumngrid

\begin{figure}[h!]
	\includegraphics[width=0.95\linewidth]{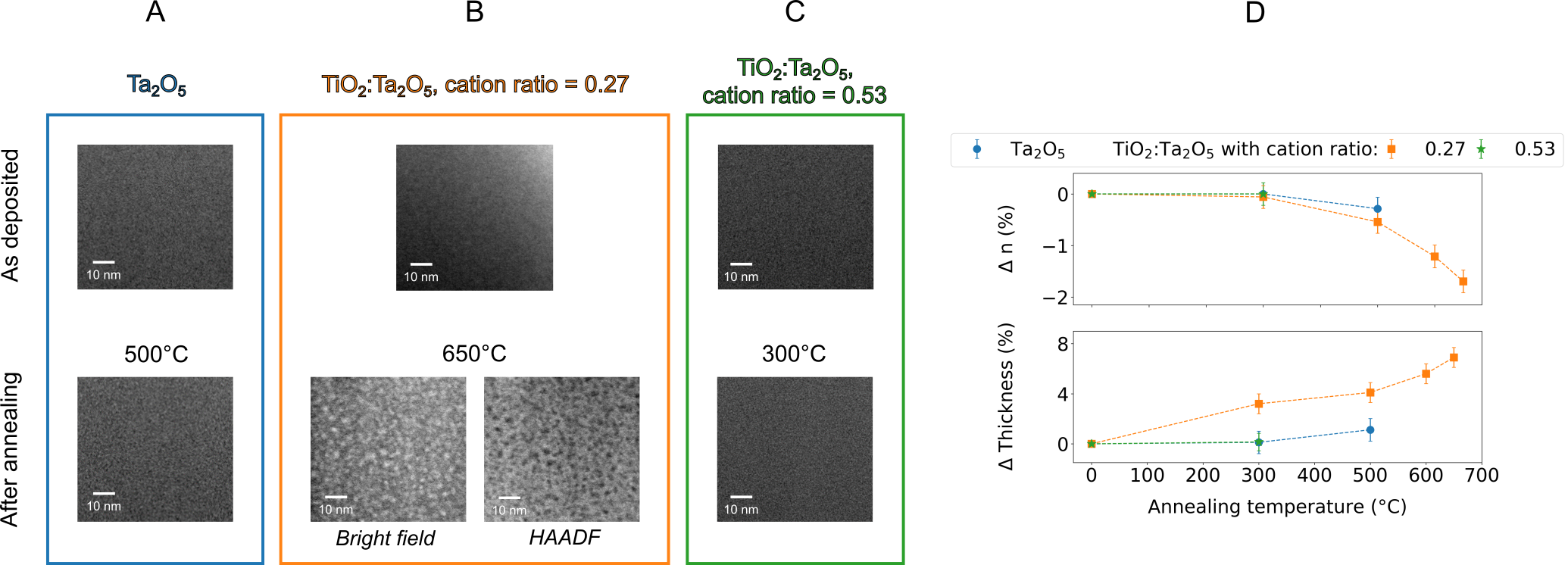}
	\caption{\label{fig:tem} TEM images and variations of the refractive index and thickness as a function of the annealing temperature. The images correspond to (A) tantala film, (B) titania-doped tantala film with a cation ratio = 0.27 and (C) titania-doped tantala film with a cation ratio = 0.53 as deposited and after annealing. All TEM images are bright field except specified. (D) Relative variation of refractive index at $\lambda$ = 1064 nm and thickness of the coatings for different annealing temperatures before the onset of crystallization.}
\end{figure}

\twocolumngrid

Annealing of the titania-doped tantala films brings further structural modifications. Figure \ref{fig:tem} shows transmission electron microscopy (TEM) images of the cross section of tantala and titania-doped tantala films as deposited and after annealing at their corresponding highest temperatures before the onset of crystallization. As expected, the as-deposited films present a dense morphology. After annealing both the tantala film and the mixed film with a cation ratio of 0.53 remain compact with no changes in the morphology within the resolution of the TEM images. However, the titania-doped tantala film with 0.27 cation concentration shows the presence of closed bubbles with a characteristic diameter of 1 – 2 nm and a surface density around 0.02 bubbles/nm$^2$, which are identified as dark regions in the high-angle annular dark-field scanning transmission electron microscopy (STEM/HAADF) image. These bubbles lead to a reduction of the packing density in the film. This is supported by the fact that the refractive index decreases and the film thickness increases with annealing temperature as shown in figure \ref{fig:tem}D. The refractive index at $\lambda$ = 1064 nm wavelength is reduced by almost 2\% between the as deposited and annealed films and is accompanied by an increase in thickness of 7\%. In contrast, both the tantala and titania-doped tantala with 0.53 cation ratio feature no significant changes in the refractive index or thickness after annealing. An analysis of the refractive index by means of the Wiener bounds approach \cite{stenzel2015physics} (Supplementary Material S3) indicates that the packing density is reduced to (0.97 $\pm$ 0.02) while a reduction in the refractive index is expected only for bubbles larger than 1 nm, which is consistent with the bubbles dimensions determined from the TEM images of figure \ref{fig:tem}B. Bubble formation induced by annealing has been reported for other sputtered oxide thin films \cite{netterfield2005low,tilsch1997effects,brown2004center,waldorf1993optical} but not for mixed films. In this case, the dopant cation ratio appears to be a key contributing factor to bubble formation. Even at lower annealing temperatures, the mixed film with 0.27 cation ratio already exhibits a clear increase in thickness in contrast with the tantala film and the mixed film with 0.53 cation ratio.

\begin{figure}[h!]
	\includegraphics[width=\linewidth]{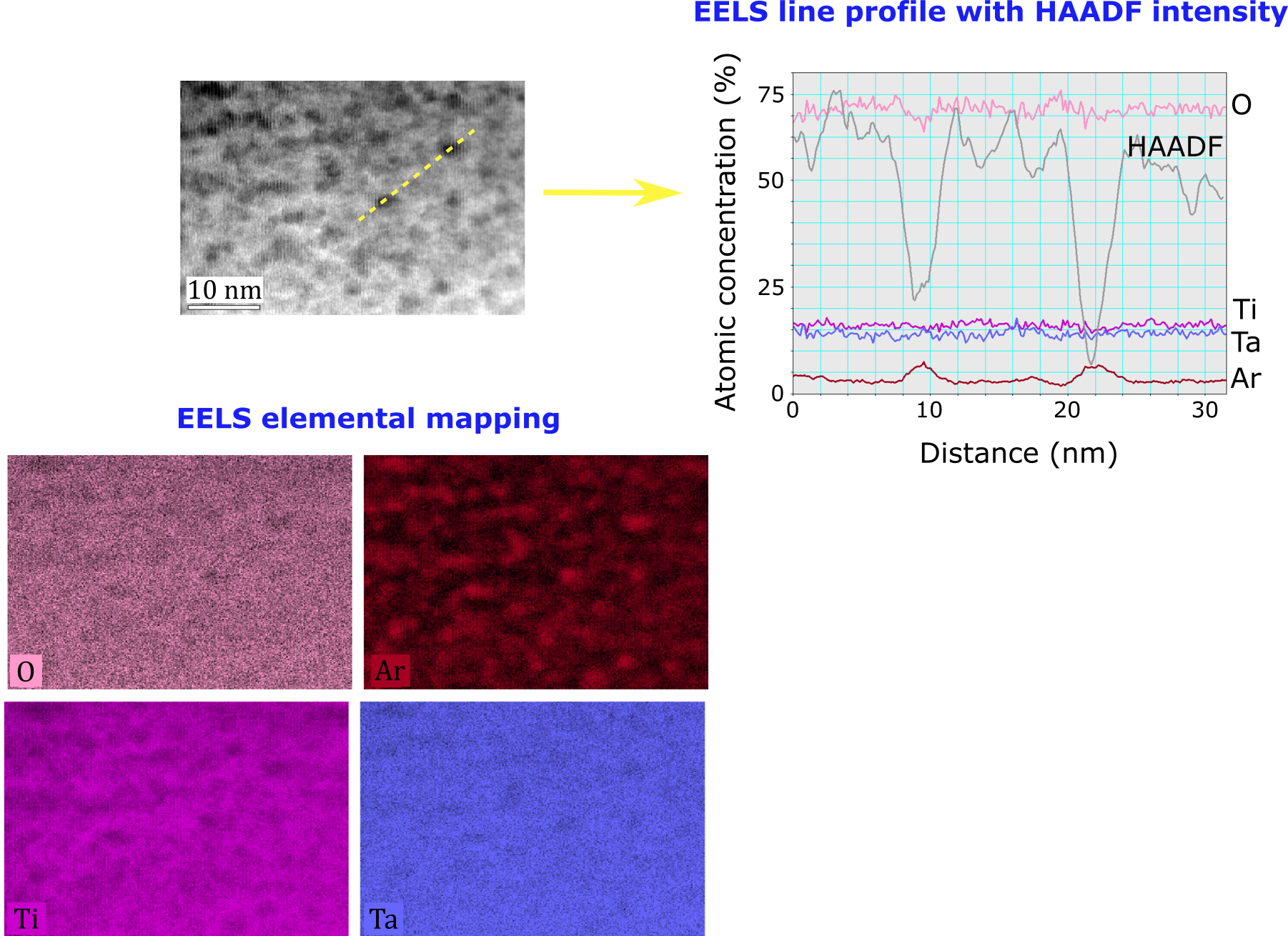}
	\caption{\label{fig:eels} TEM/EELS compositional maps and line profile of a titania-doped tantala film with a cation ratio = 0.27 after annealing imaged by TEM/HAADF. Atomic concentrations are not accurate due to the lack of standards of EELS.}
\end{figure}

Electron energy loss spectroscopy (EELS) analysis for a titania-doped tantala film with a cation ratio = 0.27 after annealing show the bubbles are mostly filled with Ar, as can be observed in figure \ref{fig:eels}. The compositional maps have Ar-rich regions that roughly correspond to dark regions in the HAADF/STEM image. This is verified by the EELS spectra taken from a line profile across the surface, which show a distinct increase only in Ar concentration that correlates with the lowest intensity regions in the HAADF/STEM image. The atomic concentrations obtained by EELS should only be considered qualitatively as this is a standardless technique and for quantification careful calibration should be carried out. From previous RBS studies done in our group, IBS films typically feature 3 - 5\% of Ar which is in good agreement with the detection of Ar by EELS in the film regardless of the presence of bubbles. During film deposition, Ar at a partial pressure around 4 $\times$ 10$^{-6}$ Torr is used to operate the ion source. The entrapped Ar in the film can be due to chemisorbed Ar or implanted Ar reflected off the target. The elevated Ar concentration in the bubbles could be due to aggregation induced by the annealing process.

\begin{table*}[!ht]
	\setlength\tabcolsep{7.5pt} % set default intercolumn whitespace width
	\begin{tabular}{cccccc}
		\hline
		\multirow{2}{*}{Film} & \multirow{2}{*}{n at $\lambda$ =  1064 nm} & \multirow{2}{*}{Optical band gap (eV)} & \multicolumn{2}{c}{Absorption loss at $\lambda$ =  1064 nm} \\
		\cline{4-5}
		& & & As deposited & Annealed \\
		\hline
		Ta$_2$O$_5$ & 2.12 $\pm$ 0.01 & 4.12 $\pm$ 0.06 & 5.7 $\pm$ 0.3 & 5.0 $\pm$ 0.2\\
		Ta$_2$O$_5$:TiO$_2$, cation ratio = 0.27 & 2.22  $\pm$ 0.01 & 3.47 $\pm$ 0.04 & 8.1 $\pm$ 0.3 & 3.9 $\pm$ 0.3\\
		Ta$_2$O$_5$:TiO$_2$, cation ratio = 0.53 & 2.30 $\pm$ 0.01 & 3.39 $\pm$ 0.04 & 79.5 $\pm$ 0.6 & 18.6 $\pm$ 0.4\\
		TiO$_2$ & 2.59 $\pm$ 0.01 & 3.28 $\pm$ 0.05 & - & -\\
		\hline
	\end{tabular}
	\caption{Refractive index at $\lambda$ =  1064 nm and optical band gap estimation for the as deposited films and absorption loss at $\lambda$ =  1064 nm for as deposited and annealed films normalized to a thickness of 250 nm. In all cases the extinction coefficient remains below the ellipsometric resolution of 1 $\times$ 10$^{-3}$ for all wavelengths $>$ 350 nm. Values for the absorption loss of titania film which is crystalline as deposited are excluded.}
	\label{table:optical}
\end{table*}

The effect of doping and annealing on the optical properties of the films was characterized. Table \ref{table:optical} summarizes the values of the refractive index at $\lambda$ = 1064 nm, the optical band gap and absorption loss at $\lambda$ = 1064 nm normalized to a thickness of 250 nm for the as-deposited films and after annealing at the highest temperature before crystallization. For tantala the refractive index and band gap agree well with reported values for films grown by reactive sputtering \cite{ngaruiya2003preparation}. The refractive indices of the mixed films follow the expected scaling with Ti doping. The absorption loss at $\lambda$ = 1064 nm for tantala and titania-doped tantala with a cation ratio of 0.27 is in the parts per million for as-deposited films and reduces upon annealing. This results in part from reduction of the concentration of oxygen defects in the thin films due to the annealing process \cite{kim2013effects,ni2008oxygen}. In particular, for the tantala film, this decrease in absorption can be associated with an increase of around 8\% in the lattice oxygen proportion as determined by XPS (Supplementary Information S2). In fact, in the current state-of-the-art coatings for gravitational wave interferometers, the absorption loss was reduced from 0.7 ppm to 0.25 ppm when the tantala layers of the mirrors were replaced by titania-doped tantala \cite{gras2017audio}.

Dopant cation ratio and annealing also have a profound impact in the mechanical loss of the coatings. The mechanical loss of a material is given by the loss angle ($\phi$), the ratio of the imaginary to real parts of the Young’s modulus, which is characteristic of the frequency dependent mechanical damping. The loss angle relates to the dissipated power as a fraction 2$\pi\phi$ of the stored energy in the test mass is being dissipated during each cycle \cite{levin1998internal}. 

\begin{figure}[h!]
	\includegraphics[width=0.9\linewidth]{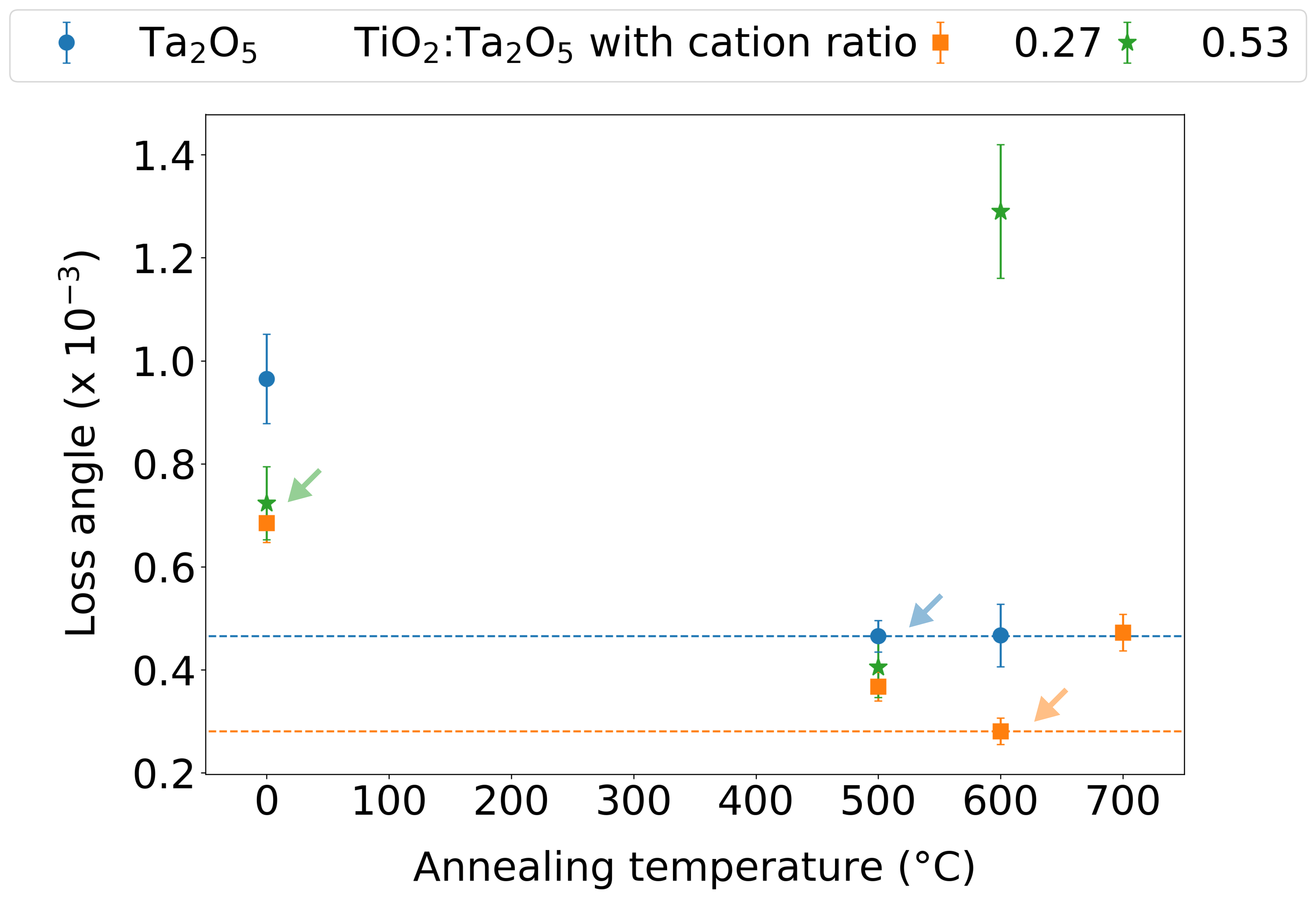}
	\caption{\label{fig:loss} Coating loss angle at 1 kHz for a tantala film and titania-doped tantala films with dopant cation ratio of 0.27 and 0.53 as a function of the annealing temperature.}
\end{figure}

The coating loss angle at a frequency around 1 kHz as a function of annealing temperature for tantala and titania-doped tantala films is presented in figure \ref{fig:loss}. For an as-deposited tantala coating the loss angle is (9.7 $\pm$ 0.9) $\times$ 10$^{-4}$ and reduces to (4.7 $\pm$ 0.3) $\times$ 10$^{-4}$ after annealing at 500\C{}, in good agreement with reported values \cite{harry2006titania,amato2018high}. The loss angle for the as deposited titania-doped tantala films is lower than for tantala, around 6.9 – 7.2 $\times$ 10$^{-4}$. After annealing at 500\C{} the loss angle for both doped films is decreased and remains lower than that of tantala. However, after annealing at 600\C{}, the loss angle of the film with a cation ratio of 0.53 increases to (13 $\pm$ 1) $\times$ 10$^{-4}$. In contrast, the titania-doped tantala film with a cation ratio of 0.27 shows a marked decrease reaching a value of (2.8 $\pm$ 0.3) $\times$ 10$^{-4}$ which is roughly a reduction of 40\% in the loss angle compared to the annealed tantala film. This reduction is comparable to previously reported values for titania-doped tantala with a dopant concentration around 20\% grown by ion beam sputtering (IBS) \cite{harry2006titania,granata2016mechanical,granata2019amorphous,aasi2015advanced,flaminio2010study,principe2015material,gras2017audio,amato2018high}. It is relevant to point out that at 600\C{} the 0.27 cation ratio titania-doped tantala reaches lower values of the loss angle and the absorption loss at 1064 nm than any of the other films in this study.

\section{Discussion and conclusions} \label{sec:discussion}

The results of the characterization of titania-doped tantala thin films presented in the previous section provide valuable insight into the structural modifications that occur with annealing and doping. In tantala films, there is an 8\% increase in the lattice oxygen concentration upon annealing that suggests that oxygen rearrangement induced by the thermal treatment is taking place. This oxygen rearrangement occurs while the structure of the film remains amorphous. Although the changes in the absorption loss and refractive index at $\lambda$ = 1064 nm, or the band gap are minimal, the oxygen rearrangement decreases mechanical loss.

The addition of titania to tantala adds other variables that influence the properties of the materials that strongly depend on dopant concentration. For the highest 0.53 cation ratio the as-deposited films are a mixture of tantala and titania and thus a phase separated material. Beyond 500\C{} titania crystalizes in its rutile phase while tantala remains amorphous. This is consistent with previous reports that showed elevated titania concentrations can lead to phase separation in sol-gel films \cite{kim2011structure}. The mechanical loss of the as-deposited film is lower than for a tantala film but after annealing at 500\C{} it reaches a similar value. The high absorption loss at $\lambda$ = 1064 nm for the as-deposited film may indicate a lack of oxygen, which is partially compensated by annealing at atmospheric conditions, but not sufficiently to reduce the absorption loss to parts per million as in tantala.

The titania-doped tantala film with a cation ratio of 0.27 shows the most significant structural modifications with annealing. The material evolves from a mixture of tantala and titania to an atomically homogenous material after annealing up to 650\C{}.  This is evidenced by the fact that the chemical environment of the Ti atoms is no longer consistent with a titania configuration. This supports previous research that suggested that tantala and titania are mixed at the atomic scale \cite{bassiri2013correlations,kim2011structure}. Moreover, the ternary compound TiTa$_{18}$O$_{47}$ crystallizes after annealing at 700\C{}. The thermal treatment also induces the formation of Ar-rich bubbles with a characteristic diameter of 1 – 2 nm and a surface density around 0.02 bubbles/nm$^2$. This effect has not been previously reported for the mixed oxide films but was suggested as a possible cause for the measured variation in the Young’s modulus of a 25\% titania-doped tantala film \cite{abernathy2014investigation}. Similar features were reported in a tantala film grown by IBS by MacLaren et. al. \cite{maclaren2019bubble}, and by Harthcock et. al. in IBS HfO$_2$ films \cite{harthcock2019impact}. Recent simulation efforts indicate that the presence and distribution of bubbles might contribute significantly to the coating loss angle of the material \cite{jiang2020atomic}. Jiang et. al. studied zirconia doped tantala and found that annealing not only promotes homogeneity by modifying the chemical environment of the Zr atoms but also allows for a homogenous distribution of voids, resulting in decreased mechanical loss. The voids reported in that study, however, are smaller than the ones found in titania-doped tantala as the size of the simulation system is roughly a 2 nm box and thus cannot reproduce nanometer sized voids. Further studies are needed to elucidate the role of voids in the loss angle of the mixed oxide coatings.

In combination the results show that the presence of Ar-rich bubbles and the atomic mixing of titania and tantala with a cation ratio of 0.27 result in a material that has the minimum coating loss angle and absorption loss at $\lambda$ = 1064 nm. Higher Ti cation ratio leads to phase separation which in turn increases mechanical loss. Therefore a path towards identifying promising mixed amorphous oxide thin films with low mechanical loss would need to account for the thermodynamics of ternary phase formation that could promote atomic mixing. A high crystallization temperature alone might not predictive of low mechanical loss and dopant cation ratio should be kept low to avoid phase separation in the film.

\section{Experimental methods} \label{sec:experimental}

\subsection{Film deposition}

The titania-doped tantala films with cation ratio of 0, 0.27, 0.53 and tantala films were deposited by reactive biased target deposition (RBTD) \cite{zhurin2000biased, hylton2000thin} using the LANS system manufactured by 4Wave, Inc. The RBTD technique involves the use of a low energy ion source that generates an Ar ion plume that is directed at a negatively biased metallic target to sputter the target. The deposition system fits six 100 mm diameter metallic targets with a mobile shutter that exposes three targets at a time. Metallic targets are biased using an asymmetric, bipolar pulsed DC power supply. The same negative bias is applied to the exposed targets but the positive pulse width and period can be individually controlled. For mixed films the Ti and Ta targets were operated simultaneously but the individual pulse widths were varied to realize different mixture proportions of the materials while the pulse period was fixed at 100 $\mu s$. The O$_2$ flow was optimized for each deposition condition in order to optimize optical properties. The base pressure was 1 $\times$ 10$^{-7}$ Torr while the process pressure was around 6 $\times$ 10$^{-4}$ Torr. The pulse width at which the Ta/Ti targets were biased to achieve the desired composition, and deposition conditions evaluated in this work are presented in Table \ref{table:dep-parameters}. The deposition rates of the films ranged from 0.004 - 0.02 nm/s. The dopant cation ratio, defined as the ratio between the titanium concentration and the sum of the titanium and tantalum concentrations (Ti/(Ti+Ta)), was obtained from the analysis of XPS spectra.   

\begin{table*}[!ht]
	\setlength\tabcolsep{5pt}
	\begin{tabular}{cccccc}
		\hline
		Film & O$_2$ flow (sccm)& Target & Pulse width ($\mu$s) & Deposition rate (nm/s) & Dopant cation ratio\\
		\hline
		I & 14 & Ta & 2 & 0.02091 $\pm$ 0.00005 & 0\\
		II & 12 & Ti/Ta & 2 / 53 & 0.01603 $\pm$ 0.00005 & 0.27 $\pm$ 0.04\\
		III & 8 & Ti/Ta & 2 / 82 & 0.00731  $\pm$ 0.00003 & 0.53 $\pm$ 0.06\\
		IV & 12 & Ti & 2 & 0.00397 $\pm$ 0.00001 & 1\\
		\hline
	\end{tabular}
	\caption{Deposition conditions and dopant cation concentration ratios obtained from XPS for the films evaluated in this work. In all cases, the pulse repetition rate was 100 $\mu s$.}
	\label{table:dep-parameters}
\end{table*}

Film thickness was kept around 200 - 250 nm for tantala and titania-doped tantala films and around 120 nm for titania films. Coatings were grown on Si (100) wafers for XPS measurements and on fused silica substrates of 25.4 mm diameter and 6.35 mm thick for the rest of the structural and optical characterization techniques employed. For mechanical loss measurements 75 mm diameter and 1 mm thick fused silica substrates were employed. Post-deposition annealing in air was carried out using a Fisher Scientific Isotemp programmable muffle furnace in which the temperature was increased at a rate of 100\C{} per hour  until a soaking temperature of 300\C{}, 500\C{}, 600\C{}, 650\C{}, 700\C{} and 750\C{} was reached. For each temperature a soaking time of 10 hours was used.

\subsection{GIXRD measurements}

GIXRD measurements were carried out on a Bruker D8 Discover Series I diffractometer with a Cu K$\alpha$ source. The incident angle was kept at 0.5$^{\circ}$  and 2$\theta$ was varied between 10$^{\circ}$ and 80$^{\circ}$. In this configuration the x-ray intensity is 90\% attenuated in a length of 340 nm of tantala and 720 nm of titania. Results for tantala and titania films are described in the Supplementary Material section. 

\subsection{XPS measurements}

XPS measurements were performed using a Physical Electronics PE 5800 ESCA/ASE system equipped with a monochromatic Al K$\alpha$ x-ray source. The photoelectron take-off angle was fixed at 45$^{\circ}$ and a charge neutralizer was used with a current of 10 $\mu A$ for all measurements. The instrument base pressure was around $1\times10^{-9}$ Torr. Spectra were analyzed using CasaXPS software (version 2.3.19) \cite{casaxps}. An in-depth analysis of XPS data is described in the Supplementary Material.

\subsection{TEM and EELS measurements}

TEM and EELS measurements were performed by Eurofins Materials Science. The samples were prepared using the in situ focused ion beam (FIB) lift out technique on a FEI Strata 400 Dual Beam FIB/SEM. The samples were imaged with a FEI Tecnai TF-20 FEG/TEM operated at 200 kV in bright-field TEM mode and high-angle annular dark-field STEM mode.

\subsection{Optical measurements}

Optical characterization of the films was realized by spectroscopic ellipsometry along with measurements of absorption loss at $\lambda$ = 1064 nm and transmittance. Ellipsometric data were collected at an angle of incidence of 60$^{\circ}$ using a Horiba UVISEL ellipsometer in a spectral range of 0.59 eV to 6.5 eV. The fitting of spectroscopic ellipsometry data was performed using the DeltaPsi2 software and several dispersion models for the materials were evaluated to obtain film thickness, refractive index and extinction coefficient. From the dispersion of the extinction coefficient, the energy band gap was estimated by the Tauc method \cite{tauc1966optical} for indirect transitions and further verified by the Cody method \cite{cody1982optical} when the range of linear dependency could not be determined unambiguously from the Tauc plot. The Supplementary Material section describes the analysis of these results. Absorption loss was measured at $\lambda$ = 1064 nm by photothermal common path interferometry \cite{alexandrovski2009photothermal}. For each sample five spots were measured in the surface in a 4 mm $\times$ 4 mm area.

\subsection{Mechanical loss measurements}

Mechanical loss measurements were performed at the LIGO Laboratory (Caltech) using a gentle nodal suspension system \cite{cesarini2009gentle, vajente2017high} for as deposited films and after annealing at different temperatures. In all cases, various resonant modes were measured in a frequency range of 1 - 22 kHz.

	\begin{acknowledgments}
This work was supported by LSC Center for Coatings Research NSF award No. 1710957.
\end{acknowledgments}

\bibliography{paper_lans}

\end{document}